\documentclass[aps,prl,twocolumn,showpacs,superscriptaddress]{revtex4-2}
\usepackage{amsmath,amssymb}
\usepackage{hyperref}
\usepackage{xcolor}
\usepackage{soul}
\usepackage[colorinlistoftodos,textsize=small]{todonotes}

\newcommand{\projA}{P_A^{\uparrow}}
\begin{document}

\title{Nonlinear Quantum Mechanics and Artificial Intelligence}

\author{Jonathan Oppenheim}
\affiliation{Department of Physics and Astronomy, University College London, London WC1E 6BT, United Kingdom}

\date{\today}

\begin{abstract}
We examine a criterion for relativistic covariance of nonlinear quantum field theory recently proposed by GPT-5 and published in Physics Letters B.
We show that this criterion inadvertently tests a different property---locality of the Hamiltonian---and is insensitive to whether the theory is nonlinear.
We recall the correct criterion, identified by Gisin and Polchinski thirty-five years ago, and reformulate their result in field-theoretic language.
\end{abstract}

\maketitle

As part of a broader effort to demonstrate AI's potential in scientific research~\cite{openai2025science}, OpenAI executives have pointed to a recently published paper by Hsu~\cite{hsu2025relativistic} as evidence that AI can contribute original ideas to physics~\cite{brockman2025,chen2025}. Hsu credits GPT-5 with proposing the core idea of the paper de novo, possibly the first published physics article where the main idea came from an LLM, and discusses the methodology in a companion piece~\cite{hsu2025ai}. We examine whether GPT-5's criterion is correct.

We show that the derived criterion, which applies Tomonaga--Schwinger (TS) integrability conditions to try to identify Lorentz-violating nonlinear modifications of quantum mechanics, instead tests a different property: whether the Hamiltonian is built from local operator densities. This is distinct from the question of linearity in the quantum state. The actual obstruction to relativistically covariant nonlinear quantum mechanics was identified by Gisin~\cite{gisin1990weinberg} and Polchinski~\cite{polchinski1991weinberg} (GP) thirty-five years ago, and is not captured by the proposed test.

\noindent\textbf{Locality versus nonlinearity.} Consider three types of Hamiltonian:
\begin{align}
H_1 &= \int d^3x\, \mathcal{O}(x), \label{eq:H1}\\
H_2 &= \int d^3x\, d^3y\, \mathcal{O}(y)\mathcal{P}(x), \label{eq:H2}\\
H_3 &= \int d^4x\, d^4y\, G_R(x,y)\mathcal{O}(y)\mathcal{P}(x), \label{eq:H3}
\end{align}
where $\mathcal{O}(x)$ and $\mathcal{P}(x)$ are local operator densities and $G_R(x,y)$ is the retarded Green's function. The first is the standard form in local quantum field theory; the second and third are nonlocal. Crucially, all three do not depend on the quantum state, and lead to dynamics which is linear in the quantum state.

In conversation with Hsu, GPT-5 proposes the Tomonaga-Schwinger (TS) integrability criterion 
\begin{align}
	[\delta/\delta\sigma(x), \delta/\delta\sigma(y)]|\Psi\rangle = 0\end{align} 
for spacelike-separated $x$ and $y$, where $|\Psi\rangle$ is the wavefunction on spacelike hypersurface $\sigma$ and $\delta/\delta\sigma(x)$ generates normal deformations of $\sigma$. For linear theories, this reduces to the microcausality condition $[\mathcal{H}(x), \mathcal{H}(y)] = 0$. Hamiltonian $H_1$ satisfies this; $H_2$ and $H_3$ do not. 

Now consider the nonlinear (state-dependent) modifications considered in \cite{hsu2025relativistic} as Eqns 7, 18 and 20 respectively:
\begin{align}
H_1' &= \int d^3x\, \langle\mathcal{O}(x)\rangle \mathcal{O}(x), \label{eq:H1p}\\
H_2' &= \int d^3x\, d^3y\, \langle\mathcal{O}(y)\rangle \mathcal{P}(x), \label{eq:H2p}\\
H_3' &= \int d^4x\, d^4y\, G_R(x,y)\langle\mathcal{O}(y)\rangle \mathcal{P}(x). \label{eq:H3p}
\end{align}
Here $\langle\mathcal{O}(x)\rangle = \langle\Psi|\mathcal{O}(x)|\Psi\rangle$ denotes the expectation value in the current state, making the Hamiltonian state-dependent and the resulting dynamics nonlinear. (More general functions of the state, such as $f(\langle\mathcal{O}\rangle)$ or functionals of $|\Psi\rangle$, do not change the analysis.) As computed in ~\cite{hsu2025relativistic}, GPT-5's criterion finds that $H_1'$, the nonlinear modification proposed by Weinberg~\cite{weinberg1989testing}, satisfies the generalized TS integrability condition, while $H_2'$ and $H_3'$ (the Kaplan--Rajendran form~\cite{kaplan2022causal}) do not.

The pattern is clear: the criterion accepts local Hamiltonians and rejects nonlocal ones, regardless of whether they are linear or nonlinear. But $H_1'$ violates Lorentz invariance via Gisin--Polchinski, so the criterion fails as a test for  nonlinear theories with local Hamiltonians which is the original theory proposed by Weinberg.

\noindent\textbf{The Gisin--Polchinski theorem.} That GPT-5's criterion is insensitive to nonlinearity can be seen from the fact that entanglement plays no role in the analysis. Yet entanglement is precisely what makes nonlinear quantum mechanics incompatible with relativity. Consider Alice and Bob sharing a singlet state $|\Psi^-\rangle = (|{\uparrow\downarrow}\rangle - |{\downarrow\uparrow}\rangle)/\sqrt{2}$. Let $Z_B$ denote Bob's Pauli-$Z$ operator and $\projA := |{\uparrow}\rangle\langle{\uparrow}|_A$ Alice's projector onto spin-up. As operators, these commute:
\begin{equation}
[\projA, Z_B] = 0.
\end{equation}
Now suppose Bob's Hamiltonian contains a Weinberg-type  term $H_B = \lambda\langle Z_B\rangle Z_B$. For the singlet, $\langle\Psi^-|Z_B|\Psi^-\rangle = 0$, and $\langle\Psi^-|H_B|\Psi^-\rangle $ vanishes. But if $\projA$ acts first, Bob's state becomes $|{\downarrow}\rangle$ with $\langle{\downarrow}|Z_B|{\downarrow}\rangle = -1$, and $H_B$ is nonzero. Alice's local operation instantaneously changes Bob's Hamiltonian. Symbolically we can apply operators and expectation values from right to left:
\begin{equation}
[\projA,H_B]|\Psi^-\rangle = \frac{\lambda}{2}|\uparrow\downarrow\rangle
\end{equation}
even though $[\projA,Z_B] = 0$. The expectation value functional does not commute with distant operations on entangled states. The requirement to consider entangled states and operators such as $\projA$ which are not proportional to a unitary, are missing from GPT-5's criterion. 

We can expand this to the context of field theory for $H_1'$. The Hamiltonian density $\mathcal{H}'(x) = \langle\mathcal{O}(x)\rangle\mathcal{O}(x)$ depends on the state through the expectation value. For field operators at spacelike separation, microcausality gives $[\mathcal{O}(x), \mathcal{O}(y)] = 0$ for $x \sim y$. But for entangled field states:
\begin{equation}
[ \mathcal{P}(x),\langle\mathcal{O}(y)\rangle\mathcal{O}(y)]|\Psi\rangle \neq 0.
\end{equation}
where  $\mathcal{P}(x)$ is a local operator which doesn't preserve the identity. This is the Gisin--Polchinski theorem: deterministic nonlinear evolution of entangled states violate Lorentz invariance because local operations instantaneously affect the distant Hamiltonian through expectation value. The TS integrability condition, which examines operator commutators $[\mathcal{H}(x), \mathcal{H}(y)]$, is blind to this. While it is recognised in ~\cite{hsu2025relativistic} that GPT-5's criterion is not enough, the claim -- that a breakdown of microcausality at later times should be considered -- is unwarranted. The missing ingredient is entanglement and projection.

Kaplan and Rajendran~\cite{kaplan2022causal} attempt to evade the GP theorem by using retarded propagators, claiming factorized evolution $U_A \otimes U_B$. But $U_A$ and $U_B$ are state-dependent, so GP still applies and the model violates both criteria.

However, the failure of nonlinear quantum mechanics runs deeper than signaling. In standard quantum theory, the density matrix is sufficient to determine measurement outcomes. In nonlinear quantum mechanics, how one decomposes the density matrix into pure states---and whether it represents a \emph{proper mixture} (classical ignorance) or an \emph{improper mixture} (reduced state of entanglement)---affects the dynamics. This is not a modification of quantum dynamics; it is a radically different theory entirely. In this regard, Kent~\cite{kent2005nonlinearity} has explored whether nonlinearity can be made consistent with no-signaling by modifying how reduced states are defined. 

\noindent\textbf{Discussion.} LLMs can assist researchers with literature searches, formatting, and initial critiques, but frequently generate plausible-sounding text that is incorrect.
In this case, the LLM proposed a criterion for relativistic covariance of nonlinear quantum mechanics that addresses a different question than intended. The actual obstruction---that expectation values fail to commute with distant operations on entangled states---was identified 35 years ago by Gisin and Polchinski in highly cited publications, that would have been accessible to GPT-5, making the error more notable.  

Hsu deserves credit for his transparency about the LLMs contribution and for his thoughtful reflections on the difficulties of verification~\cite{hsu2025ai}. This transparency is precisely what makes the case instructive: it illustrates a characteristic failure mode in which LLMs generate sophisticated-looking analysis that misses the conceptual core of a problem. If plausible but incorrect AI-generated reasoning can appear convincing even under the scrutiny of distinguished experts, the broader scientific community should be extremely cautious. As AI tools become more prevalent in research, rigorous verification becomes more essential~\cite{oppenheim2025slop}. Notably, no current frontier models---GPT-5.1, Claude Opus 4.5, or Gemini 3 pro---identified the error when asked to review~\cite{hsu2025relativistic}; all required explicit prompting to recognize the flaw, at which point GPT-5.1 stated that the analysis ''landed on the crux of the issue'' and that the ''challenge was exactly the right one''.

While LLMs are also useful tools that researchers can use to critically interrogate their work, the sycophancy of LLMs poses a particular danger. Models trained to please users and optimize engagement are the opposite of what science needs: adversarial critics who point out flaws rather than polish prose. Science requires tools that challenge assumptions, not reinforce them. As AI-generated content becomes more ubiquitous and more sophisticated, the scientific literature faces a challenge: a rising tide of plausible-sounding but flawed analysis that overwhelms traditional peer review. The challenge is to develop the filters during the flood.

\noindent\textbf{Acknowledgements:} Claude Opus 4.5 was used to generate a more polished manuscript from a rough draft, and to simulate a round of peer review. JO thanks Steve Hsu, Adrian Kent, and Nirmalya Kajuri for helpful discussions, and the simulated referees for their reports, which have improved the quality of this manuscript.

\bibliography{refs}

\end{document}